\documentclass[12pt,a4paper]{article}
\usepackage[utf8]{inputenc}
\usepackage[T1]{fontenc}

\usepackage{amsmath}
\usepackage{amssymb}
\usepackage{tabularx}
\usepackage{authblk}
\usepackage{graphicx}
\usepackage{color} 
\usepackage[linktoc=page]{hyperref}
\hypersetup{
colorlinks=true,
linkcolor=blue,
bookmarks=true,
citecolor=[rgb]{0.54,0,0}
}
\author{Yoav Zigdon}
\affil{{\normalsize \textit{Department of Applied Mathematics and Theoretical Physics, University of Cambridge, Wilberforce
			Road, Cambridge CB3 0WA, United Kingdom}} \\
	{\normalsize yz910(at)cam.ac.uk}}
\date{}

\title{Stringy Forces in the Black Hole Interior}
\begin{document}
	\maketitle
		\begin{abstract}
		Effective field theories break down inside large black holes on macroscopic scales when tidal forces are string-sized. If $r_0$ is the horizon radius and $\alpha'$ is the square of the string scale, the 4D Schwarzschild interior is strongly curved at $ \big(r_0 \alpha' \big)^{1/3}$. Infalling massless probes that reach this scale stretch and become excited strings. I generalize this picture for a wide class of black hole solutions in string theory. For the black hole dual to the large-$N$ BFSS model in a thermal state, and denoting $\ell_P$ the Planck length, tidal forces are stringy at $r_0 \left(\frac{r_0}{N^{1/3} \ell_P}\right)^{3/11}$, which is greater than the scale where string perturbation theory breaks down for sufficiently large $r_0/\ell_P$. For 4D Kerr, there is a range of spin parameters for which the inner horizon is to the future of the scale of stringy curvature. These results specify the portion of black hole interior solutions where effective field theory can be used; beyond these scales, one must resort to other methods. 
	\end{abstract}
	\newpage 
	\tableofcontents
	\section{Introduction} 
	It is often assumed that the physical picture of black hole interiors is the vacuum until the singularity or a core of size comparable to the Planck scale $\ell_P$. A problematic aspect of this description is the contradiction between the principle of quantum unitary evolution and Hawking's calculation of the state of radiation emitted from black holes \cite{Hawking75}.
	
	It has been argued that the assumptions of a finite-dimensional Hilbert space with unitary S-matrix describing black hole microstates imply that effective field theory near the horizon breaks down  \footnote{The literature contains papers that argued that effective field theory breaks down outside 4D Schwarzschild black holes at proper distances $(\ell_P ^2 r_0)^{\frac{1}{3}}$~\cite{Casher:1996ct}-\cite{Marolf:2003bb} and $\sqrt{\ell_P r_0}$~\cite{Bousso:2023kdj}-\cite{Banks:2024imv} (see also \cite{Ong:2023lbr}) away from the horizon  where $r_0$ is the horizon radius.} \cite{Itzhaki96},\cite{Mathur09},\cite{Polchinski12}. The latter reference has introduced the notion of a ``firewall'' at the horizon where a structure with high energy is localized, in the vicinity of which infalling observers die. A physical picture that also has structure at the horizon scale follows from the ``fuzzball'' idea , which was reviewed in \cite{Fuzzball}, that the black hole interior is supplanted by horizon-free and singularity-free bound states of string theory whose wavefunctions are supported on the horizon scale. On the other hand, trusting effective field theory near and behind the horizon, and including wormholes in the gravitational path integral, have given rise to a picture of the black hole interior connected to Hawking radiation by Einstein-Rosen bridges \cite{Maldacena13}, as well as Page curves consistent with unitary evolution and the finiteness of the Hilbert space of black hole microstates \cite{Penington:2019kki},\cite{Almheiri:2019qdq}. 
	\\ This paper aims to constrain the use of effective field theories behind the horizon where string theory effects become important when tidal forces on infalling probes are ``string-sized''. This paper starts with the premise that effective field theory is valid on the macroscopic (outer) horizon scale because of small string coupling and small curvature measured in string units and is motivated by specifying which region in the interior admits a good effective field theory description.

	In weakly coupled string theory, there is a scale much larger than $\ell_P$, where one expects significant modifications to the General Relativity (GR) picture due to higher-curvature terms in the effective action, namely the string scale $\ell_s=\sqrt{\alpha'}$. However, there is yet a much greater scale, $r_* = (\sqrt{12}\alpha' r_0 )^{\frac{1}{3}}$,
	for 4D Schwarzschild for example \cite{Emil94}, where $r_0$ is the scale of the horizon, where string theory modifies the GR picture. The string scale determines tidal forces on the infalling matter at that macroscopic scale, as shown in the next section by computing curvature invariants.
	
	If a low-energy infalling massless mode which belongs to the spectrum of string theory (e.g.~a graviton) happens to reach $r_*$, its kinetic energy would be converted to excitations of the fundamental string; these excitations then stretch as they dive inside and are trapped within $r_*$ due to gravity. The technical reason for the energy conversion is that when writing an approximate CFT$_2$ describing string propagation in the black hole interior, there are time-dependent interaction terms that couple the massless mode in question with high-frequency modes of the fundamental string, which become significant at $r=r_*$. Moreover, higher-derivative terms in the effective action constructed solely from the Riemann tensor are all of the same order at $r=r_*$, making the genus-zero worldsheet theory strongly coupled.
		
	A few connections to the literature are written. Reference \cite{Horowitz90} pointed out that tidal forces excite test strings when they propagate in wave-fronted backgrounds and calculated occupation numbers of string modes produced during the propagation.  The papers \cite{Myers96},\cite{Ross97} discussed large tidal forces in the vicinity of (would-be) horizons of certain black strings, while \cite{Maeda99} did so for a Vaidya solution near an inner horizon. Reference \cite{Silverstein12}  described tidal forces creating excitations of the string near a null singularity and argued that interactions with D0-branes limit them. The paper \cite{Arnold:2012qg} estimated the stopping distance of certain high energy excitations in $\mathcal{N}=4$ super-Yang Mills by computing the lengths of bulk geodesics describing infalling gravitons that transition to large classical string loops near a black brane horizon. Reference \cite{Silverstein14} estimated the number of open strings between infalling D-particles in the Schwarzschild interior. \\
	More recently, reference \cite{Emil20} considered an $\frac{1}{8}$-BPS background carrying D1-D5-P charges that admits string-sized and compactification-sized tidal forces which excite, stretch and trap test strings that propagate toward the cap of the geometry (see also \cite{Tyukov:2017uig}-\cite{Guo:2024pvv}). Reference \cite{Dodelson:2020lal} considered light-like singularities in bulk-to-bulk two-point functions in the 5D AdS Schwarzschild black hole exterior and their resolutions due to tidal forces that lead to particle creation on the worldsheet. The papers \cite{Horowitz22},\cite{Horowitz23},\cite{Horowitz24} considered extremal black holes in asymptotically AdS, extremal Kerr, and near-extremal Kerr-Newman black holes in asymptotically flat spacetime, finding that effective field theory breaks down near and outside the horizon due to large tidal forces.   
	To the author's knowledge, however, the scales written in this paper in which effective field theories of string theory inside 4D charged, rotating black holes and black p-branes break down have not appeared before. The logic applied in this paper has also been used in the Swampland program \cite{Vafa05}, \cite{Eran}, though the latter has often focused on extremal and BPS black holes.
	
	The paper is organized as follows. 	In the next section, scales are written when the Kretschmann scalar, which measures tidal forces, is determined by $\frac{1}{(\alpha')^2}$ for various black hole interiors. Starting with 4D Schwarzschild in subsection \ref{subs:Sch}, it is shown that $\alpha'$ corrections become significant at the scale of stringy tidal forces. Subsection \ref{subs:Kerr} is devoted to the 4D Kerr relevant to the Universe in which we live, and the scale of strong curvature is encountered generically in a narrow layer about the equator at the same radial coordinate $r_*$ of Schwarzschild. For the 4D Reissner-N\"ordstrom black hole, the main result in subsection \ref{subs:RN} is the existence of solutions with an inner horizon much greater than the string scale, which is nonetheless smaller than $r_*$, namely $\ell_s \ll r_- \ll r_*$. Additionally, decoupling limits of black p-brane solutions  \cite{Horowitz91},\cite{Itzhaki98} admit interior scales of stringy tidal forces, in particular the D0-brane black hole, which are listed in the subsections \ref{subs:D0}-\ref{subs:Dp} and summarized in Table~\ref{Table1}. 
	The paper ends with a summary and comments in Section \ref{subs:Summ}. The cases of the BTZ and the 2D linear dilaton black hole solutions, which admit exact worldsheet descriptions, appear in an appendix. 
	   
	\section{Examples}
	\label{subs:Examples}
	\subsection{Schwarzschild}
	\label{subs:Sch}
\begin{figure}[h!]
	
		\begin{center}
			\includegraphics[scale=1]{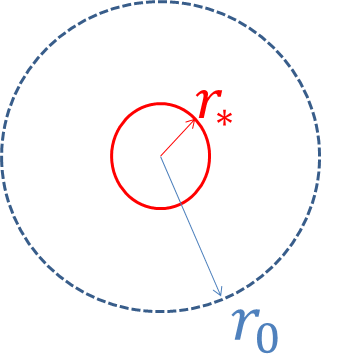}
		\end{center}
		\caption{The horizon is depicted in the dashed circle, and the solid red circle represents the spacelike surface where effective field theory breaks down, corresponding to $r_*\sim \Big( \alpha' r_0\Big)^{\frac{1}{3}} \gg \sqrt{\alpha'}$. 
			If the string scale corresponds to the radius of the nucleus of the atom, $\ell_s \approx  10^{-14} ~m$, and considering the horizon radius of Sagittarius $A*$, $r_0 \approx 1.2 \times 10^{10}~m$, the special scale reads $r_* \approx 1.6\times 10^{-6}~ m~$.
		} 
			\label{fig:Schwrtzschild}
	\end{figure}

	In the context of Type II or the heterotic string theory with a $\mathbb{T}^6$ compact manifold, the Schwarzschild line element is
	\begin{equation}
		\label{Schw} 
		ds^2 = -\frac{dr^2}{\frac{r_0}{r}-1}+\left( \frac{r_0}{r}-1\right)dt^2   + r^2 d\Omega_2 ^2+ds_{\mathbb{T}^6} ^2~.
	\end{equation}
	The Kretschmann scalar, defined as a squared of the Riemann tensor, is given by
	\begin{equation}
		K \equiv R_{\mu \nu \alpha \beta} R^{\mu \nu \alpha \beta} = \frac{12 r_0 ^2}{r^6}~.
	\end{equation}
	This quantity admits an interpretation as a measure of tidal forces on infalling matter that moves along geodesics \cite{Reall}.
	The Weyl tensor squared coincides with $K$ for Ricci-flat backgrounds and has a similar interpretation.  One can ask, when do these tidal forces become stringy? This is simply answered by equating $K$ to the scale $\frac{1}{(\alpha')^2}$:
	\begin{equation}
		\label{r*}
		K = \frac{1}{(\alpha')^2} \Rightarrow r_* = \left( \sqrt{12}r_0 \alpha'\right) ^{\frac{1}{3}}~.
	\end{equation}
	For a macroscopic black hole $r_0 \gg \sqrt{\alpha'}$ and the following hierarchy applies:
	\begin{equation}
		r_* \gg \sqrt{\alpha'}~.
	\end{equation}
	As explained below, a closely related interpretation of the scale $r=r_*$ is that this is when the black hole interior becomes strongly curved - higher curvature corrections are all of the same order of magnitude. In particular, General Relativity breaks down. 
	In the context of the heterotic string theory \cite{Emil1984},\cite{Emil1985}, a higher derivative term in the target space effective Lagrangian density is \cite{Emil1985b}
	\begin{equation}
		\label{het2} 
		L_{R^2} ^{het} = \frac{\alpha'}{8\kappa_0 ^2} e^{-2\Phi} \sqrt{-G} R_{\alpha \beta \gamma \delta} R^{\alpha \beta \gamma \delta}~. 
	\end{equation}
	In this equation, $\kappa_0$ appears in the normalization of the tree level Lagrangian, $G$ is the determinant of the string frame metric, and $\Phi$ is the dilaton. 
    Reference \cite{Gross1986} calculated the four-graviton scattering amplitude in the heterotic theory, from which one deduces  $(\alpha')^3$ terms  in the Lagrangian density
	\begin{align}
		\label{het3} 
		&L_{R^4} ^{het}= \frac{\zeta(3)(\alpha')^3}{48 \kappa_0 ^2} e^{-2\Phi} \sqrt{-G} \Big(  2 R_{abcd} R_{e ~~~f} ^{~~bc} R^{aghe} R^{f~~~~ d} _{~~gh} + R_{abcd} R_{ef } ^{~~~cd} R^{aghe} R^{f~~~~~b} _{~~gh} \Big)  ~.
	\end{align}
	Here, $\zeta(3)\approx 1.2$ is the Ap\'ery's constant, and the scheme where no explicit Ricci tensor or scalar appears has been adopted, as in reference \cite{Myers1987}.
	The ratio of the equations (\ref{het3}) and (\ref{het2}), evaluated at the Schwarzschild solution,  is
	\begin{equation}
		\frac{L_{R^4} ^{het}}{ L_{R^2} ^{het}} =\frac{\zeta(3) (\alpha')^2 r_0 ^2}{8r^6}~. 
	\end{equation}
	Thus, the quartic term $(\alpha')^3 R^4$ in the heterotic string effective action is subdominant as long as $r\gg r_*$. However, once probes reach $r_*$, they can no longer be described using an effective field theory approach.
    The target space effective Lagrangian density of Type II superstring contains $(\alpha')^3$ terms similar to Eq.~(\ref{het3}) \cite{Gross1986b},\cite{Myers1987} (see also \cite{Yiming}). Plugging the Schwarzschild solution into this quartic term, one finds $ 	L_{R^4} ^{II}=\sqrt{-G}e^{-2\Phi} \frac{9 \zeta(3) (\alpha')^3 r_0 ^4}{32 \kappa_0 ^2 r^{12} }$.
	This implies that $L_{R^4} ^{II}  \propto \frac{1}{\alpha'} $ is encountered at $r\approx (r_0 \alpha')^{\frac{1}{3}} $ and therefore the scale $r_*$ admits an interpretation as the onset of strong curvature in the black hole interior.
	See Fig.~\ref{fig:Schwrtzschild}~.
	
	Suppose one uses GR to predict what happens to a probe when it reaches $r_*$.  First, one can consider the infalling time between $r_0$ and $r_*$ as predicted by the Schwarzschild metric. For a zero angular momentum and massless probe, to leading order in $\frac{\ell_s}{r_0}$, the time is $\frac{\pi}{2}r_0 $, set by the light-crossing time.
	However, in GR, nothing special would happen to the probe at the time $r_*$.  
		
	 	 It is explained below that there is an important modification to the experience of infalling massless modes in a string theory context.
	 	 
	The phenomenon of string propagation is described by a worldsheet CFT$_2$. An approximate spacetime for $r\gtrsim  r_*$, on which an approximate CFT$_2$ can be considered, is Eq.~(\ref{Schw}).
	The bosonic action of a closed string is 
	\begin{equation}
		\label{Polyakov}
		I = -\frac{1}{4\pi \alpha'} \int d^2 \sigma \sqrt{-h} h^{ab} \partial_a X^{\mu} \partial_b X^{\nu} G_{\mu \nu}(X)+S_{\text{ghosts}}~, 
	\end{equation}
	where $G_{\mu \nu}$ is the spacetime string metric, $X^{\mu}=X^{\mu}(\sigma)$ is the embedding of a string in spacetime, $\sigma$ is a composite notation for worldsheet position, $h_{ab}$ are worldsheet metric components and $S_{\text{ghosts}}$ describes ghosts. This action was used for the Schwarzschild target spacetime by \cite{Emil93} to estimate the extent of string excitations in that geometry and the rate of stringy Hawking radiation.  Suppose one expands $X^{\mu} (\sigma)$ near a macroscopic horizon described by Kruskal coordinates in modes relative to the worldsheet time $\sigma_2$:
	\begin{equation}
		X^{\mu} (\sigma) = x_{C.M.} ^{\mu} + p^{\mu} _{C.M.} \sigma_2+R_w ^{\mu} \sigma_1  +i \sqrt{\frac{\alpha'}{2}} \sum_{ n\in \mathbb{Z}/\{0\}} \frac{1}{n} \Big( \alpha^{\mu} _n e^{-i n (\sigma_1 +\sigma_2)}+\widetilde{\alpha}_n ^{\mu} e^{in (\sigma_1 - \sigma_2)} \Big)~.
	\end{equation}
	A gauge choice can fix two fields $X^{\mu}$.
	From left to right, the interpretations of the terms on the R.H.S. are: Center of mass position, center of mass momentum, winding around possible compact dimensions, and the ones in the sum represent excited string modes. Even though the worldsheet theory has not been quantized on (\ref{Schw}), an approximate quantization on a weakly curved region leads to the realization that Eq.~(\ref{Polyakov}) includes worldsheet time-dependent interaction terms that transfer energy from the center of mass momentum mode of the infalling massless string probe to higher modes corresponding to string excitations.\\
	A quantitative question is: What are the occupation numbers of string modes created by the stringy tidal forces? Approximated calculations of these quantities were performed in the context of pp-wave backgrounds \cite{Horowitz90}, near the horizon of charged dilatonic black holes in \cite{Horowitz1997b} and for an $\frac{1}{8}$-BPS D1-D5-P background \cite{Emil20}; however, in the present system it is presently difficult to answer since an exact CFT for the Schwatrzschild black hole is unknown, implying that the $r\leq r_*$ region is inaccessible with current methods. In particular, the Penrose limit, which focuses on near-geodesics regions, is not helpful for this system.~\footnote{References \cite{Blau},\cite{Blau2004} extrapolated geodesics until the singularity, which is unreliable in a string theory context.}

	Another significant scale is associated with the volume of the compact manifold:
	\begin{equation}
		K = \frac{1}{\text{Vol}(\mathbb{T}^6) ^{\frac{2}{3}}} \Rightarrow r_{*int} = \left( \sqrt{12} r_0 \text{Vol}(\mathbb{T}^6)^{\frac{1}{3}}  \right)^{\frac{1}{3}}~.
	\end{equation}
	If $\text{Vol}(\mathbb{T}^6) \leq (\alpha')^3$, then $r_{*int}\leq r_*$,  and the effective field theory description breaks down at $r_*$.
	In the reverse case, the infalling center of mass momentum probe would be converted into momentum modes in $\mathbb{T}^6$ - which again would lead to a significant modification to the 4D GR prediction.~\footnote{If the Universe in which we live admits a dark dimension of scale $R=10^{-6}m$ \cite{Dark}, then the scale when energy of infalling probes is converted to internal modes on the extra dimension, in the Sagittarius A* black hole interior, is $(\sqrt{12} r_0 R^2 )^{\frac{1}{3}}\approx 34~cm$. } 
	\\
	Two comments are in order.  First, in $D> 4$ spacetime dimensions, the scale of strong curvature inside a Schwarzschild-Tangherlini black hole is
	\begin{equation}
		r_* (D) = \Big[(D-1)(D-2)^2 (D-3) \Big]^{\frac{1}{2(D-1)}}r_0 ^{\frac{D-3}{D-1}} \ell_s ^{\frac{2}{D-1}}~.
	\end{equation}
	In the limit of large $D$ and fixed $g_s ^2 S$, with $g_s$ the string coupling and $S$ the Bekenstein-Hawking entropy, $r_* (D)\to \sqrt{D}\ell_s$, which again exemplifies that the black hole interior enhances the naive scale where $\alpha'$ corrections are important by a large factor. 
	Second, in the context of M-theory, the semiclassical description  in a $D\leq 10$ dimensional Schwarzschild-Tangherlini interior cannot be trusted when curvature invariants are determined by the 11th-dimensional Planck scale, namely at
	\begin{equation}
		\label{r*M}
		r_* ^{(M)} (D)=\Big[(D-1)(D-2)^2 (D-3) \Big]^{\frac{1}{2(D-1)}}r_0 ^{\frac{D-3}{D-1}} \ell_P ^{\frac{2}{D-1}}~.
	\end{equation}
	\footnote{ The paper \cite{Poisson:1988wc} suggested attaching black hole interiors with an expanding cosmology, and references \cite{Frolov:1988vj},\cite{Frolov:1989pf},\cite{Poisson:1990eh} assumed that this starts at the scale $r_0 ^{\frac{1}{3}} \ell_P ^{\frac{2}{3}}$.}
	 As long as the 11th-dimensional circle is small, this provides a significant gain in the regime of validity of the description relative to the weakly curved and weakly coupled string theory description. 
	\\
	 The conclusion of this subsection is the existence of the macroscopic scale $r_*\gg \ell_s$  when effective field theories of string theory based on the (\ref{Schw}) geometry are not valid, and string modes are created from infalling massless probes. 
		
	\subsection{Kerr} 
	\label{subs:Kerr}
	This subsection presents a calculation of scales in the Kerr geometry where tidal forces are string-sized. \footnote{Reference \cite{LimaJunior:2020fhs} solved the geodesic deviation equations in the symmetry axis of the Kerr geometry. }
	The line element of the Kerr solution with mass $M$ and angular momentum $J$, multiplied by a $\mathbb{T}^6$, is given by
	\begin{align}
		ds^2 &= -\left(1-\frac{r_0 r}{r^2+ a^2 x^2}\right)dt^2 -\frac{2rr_0 a (1-x^2)}{r^2+a^2 x^2}dt d\phi + \frac{r^2+a^2 x^2}{r^2  - r_0 r +a^2}dr^2 \nonumber\\
		&~~~~~+\frac{r^2+a^2 x^2}{1-x^2}dx^2 + \left(r^2 +a^2 + \frac{r_0 r a^2}{r^2+a^2 x^2} (1-x^2)\right) (1-x^2) d\phi^2 +ds_{\mathbb{T}^6} ^2 ~,
	\end{align}
	where
	\begin{equation}
		x=\cos(\theta)~,~a=\frac{J}{M}~,~ r_0 = 2 G M~.
	\end{equation}
	The horizons are located at
	\begin{equation}
		r_{\pm}  = \frac{1}{2} \Big( r_0 \pm \sqrt{r_0 ^2 - 4a^2} \Big)~.
	\end{equation}
	The Riemann tensor squared is given by
	\begin{equation}
		\label{KKerr} 
		K= \frac{12 r_0^2}{(r^2 + a^2 x^2)^6} \left[r^2 - a^2 x^2 \right] \left((r^2 + a^2 x^2)^2 - 16 r^2 a^2 x^2 \right)~.
	\end{equation}
     	The physics at the equator is simple: There are geodesics purely at  $x=0$ of the Kerr interior on which the center of mass momentum modes of test strings can move. From Eq.~(\ref{KKerr}), one finds the same $K(r,x=0)$ as in the Schwarzschild black hole, and then  $r_* \propto (r_0 \ell_s ^2)^{\frac{1}{3}}$ is when the equator is strongly curved, for any $0\leq a\leq G M$, and the same conclusion on the transition between the center of mass momentum modes to excited string modes holds. \\
     	Can one find where the string scale determines the Kretschmann scalar outside the equator? The answer is positive. Before pinpointing to such regions, a necessary condition for their existence is derived below.  Let us define
     	\begin{equation}
     		\tilde{a} \equiv \frac{2a}{r_0}~,~\tilde{r}\equiv \frac{r}{r_0}~,~
     		z(\tilde{r},x)\equiv  \tilde{r}^2 + \frac{1}{4}\tilde{a}^2 x^2 ~,~z_* \equiv \left(\frac{12(\alpha')^2}{r_0^4}\right)^{\frac{1}{3}}~.
     	\end{equation}
     	Eq.~(\ref{KKerr}) for the Riemann tensor squared is equal to $\frac{1}{(\alpha')^2}$ provided that
     	\begin{equation}
     		\label{sextic}
     		\frac{r_0^4}{12(\alpha')^2}z^3 (z-z_*)(z^2 + z_* ^2 + zz_*) + \frac{9}{2} \tilde{a}^2 x^2 \Big(z- \frac{1}{3}\tilde{a}^2 x^2\Big)^2=0~.
     	\end{equation}
     	Since the two terms on the L.H.S of Eq.~(\ref{sextic}) are positive definite for $z>z_*$ if there are solutions, they can only occur for $z\leq z_*$. Therefore, a necessary condition for the existence of strong curvature in the Kerr interior is
     	\begin{equation}
     		\label{Necessary}
     		 \tilde{r}^2 + \frac{1}{4}\tilde{a}^2 x^2\leq \left(\frac{12(\alpha')^2}{r_0^4}\right)^{\frac{1}{3}}~. 
     	\end{equation}
     	One lesson from this inequality is that when $r\gg r_*$, no stringy tidal forces exist. An interpretation of this result is that the presence of rotation applies a repelling force on infalling probes such that gravity and the centrifugal force cannot collude to produce larger tidal forces in Kerr. Another consequence is that when the rotation parameter is $\tilde{a}\sim O(0.5)$, as in a set of black holes in our Universe \cite{Reynolds:2013qqa}, and one considers $x\sim O(0.5)$ to explore the region outside the equator, the inequality is violated - meaning that generically there are no string-sized tidal forces outside the equator. This is related to the distance between this region and the ring, timelike singularity of Kerr being sufficiently large.  A corollary is that large, near-extremal Kerr solutions do not admit strong curvature in generic points outside the equator (though the paper \cite{Horowitz23} has demonstrated that perturbations about extremal Kerr in theories with higher curvature corrections invalidate effective field theory on the horizon scale). In fact, if the inner horizon is much greater than the string scale, it follows that $\frac{\ell_s}{r_0}\ll \tilde{a}^2$ and (\ref{Necessary}) is violated for generic $x$.
     	
     	Two approximate solutions to Eq.~(\ref{sextic}) away from the equator are written below.
     	
     	1. For $r\ll a x$, one finds that 
     	\begin{equation}
     		K \approx - \frac{12r_0^2}{a^6 x^6}~,
     	\end{equation} 
     	which is of order $-\frac{1}{(\alpha')^2}$ when
     	\begin{equation}
     		\label{acritic}
     		\tilde{a} \sim  \tilde{a}_c=\Big( \frac{\ell_s}{r_0}\Big)^{\frac{2}{3}}~.
     	\end{equation}
     	This means that for a narrow window of angular momenta $\tilde{a} = \tilde{a}_c + \delta \tilde{a}$, with $\delta \tilde{a} \ll  \tilde{a}_c$,  effective field theory breaks down at $\tilde{r}\ll \tilde{a}_c$. However, since Eq.~(\ref{acritic}) implies that $a\ll r_0$, the inner horizon is sub-stringy $r_- \approx \frac{a^2}{r_0} \sim \ell_s \left(\frac{\ell_s}{r_0}\right)^{\frac{1}{3}}$. Nonetheless, a probe that reaches $r\approx \frac{a}{100}$, for instance, which is outside the inner horizon for a sufficiently large ratio of $\frac{r_0}{\ell_s}$, experiences stringy tidal forces. 
     	
     	2. For  $ax\ll r$ (near the equator or for very low angular momenta black holes), one finds that $z\approx z_*$ is an approximate solution that occurs at
     	\begin{equation}
     		r\approx r_*=(\sqrt{12} \alpha' r_0)^{\frac{1}{3}}~.
     	\end{equation}
     	For example, considering the poles $x=\pm 1$, a controlled approximation $a \ll r_*$ requires $\tilde{a}\ll \left(\frac{\ell_s}{r_0}\right)^{\frac{2}{3}}$. Such solutions have  tiny inner horizons $r_- \ll \ell_s$. Alternatively, one can consider order half rotation parameters, $\tilde{a} = O(0.5)$. Then the angular separation between the plane of rotation and the region of string-sized tidal forces is $\delta \theta \sim \left(\frac{\ell_s}{r_0}\right)^{\frac{2}{3}}$. 
     	
     		Next, it is important to show that there is a non-trivial range of values of $\tilde{a}$ for which effective field theory breaks down at $r_*>r_-$, such that the conventional cosmic censorship conjecture does not constitute a reason for the breakdown of effective field theory. A comparison between the inner horizon scale and $r_*$ is made using numerics. Choosing $\frac{r_0}{\sqrt{\alpha'}}=100$, the following density plot shows in which region in parameter space $(x,\tilde{a})$ the inner horizon surpasses the scale of string scale curvature (blue) and the complementary region (light yellow).
     	
     	\begin{figure}[h]
     		\centering 
     		\includegraphics[scale=0.5]{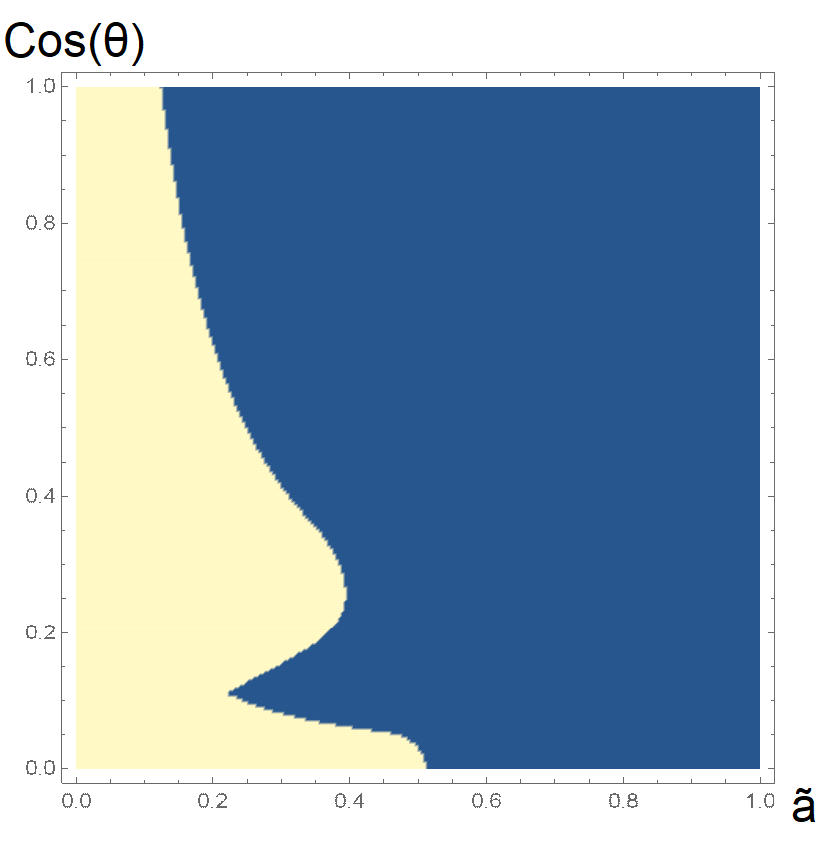}
     		\caption{ The blue region describes angles and rotation parameters for which $r_- (\tilde{a})>r_*(\theta,\tilde{a})$. The light yellow region shows for which parameters $r_- (\tilde{a})<r_*(\theta,\tilde{a})$. Typically, when the rotation axis is approached, and as $\tilde{a}$ increases, $r_->r_*(\theta,\tilde{a})$, consistent with the calculations of this subsection. The plot is for $\frac{r_0}{\sqrt{\alpha'}}=100$ and indicates that when $\tilde{a}\gtrsim 0.5$, the inner horizon is greater than $r_*(\theta,\tilde{a})$ throughout all angles. }
     	\end{figure}  
     	 
	To summarize, a family of rotating black hole solutions exhibits string-sized tidal forces on infalling massless modes that turn into excited strings before reaching the inner horizon.
	
	\subsection{Reissner-N\"ordstrom}
	\label{subs:RN}
	In the heterotic string theory, one can construct electrically charged black hole solutions with geometry
	\begin{equation}
		ds^2= -\left(1-\frac{r_0}{r} +\frac{r_Q ^2}{r^2}\right)dt^2+ \frac{dr^2}{1-\frac{r_0}{r} +\frac{r_Q ^2}{r^2}} + r^2 d\Omega_2 ^2 + ds^2 _{\mathbb{T}^6}~,
	\end{equation} 
	and $U(1)$ gauge field
	\begin{equation}
		A = \frac{Q}{r}dt~.
	\end{equation}
	The length scales of the solution are
	\begin{equation}
		r_0 = 2 GM ~,~ r_{Q} = Q \sqrt{G}~.
	\end{equation}
The horizons are at
	\begin{equation}
		r_{\pm} = \frac{r_0 \pm \sqrt{r_0 ^2 - 4r_Q ^2}}{2}~.
	\end{equation}
	Reference \cite{Crispino:2016pnv} calculated tidal forces in this black hole in the context of GR.
 The Kretschmann scalar is given by
	\begin{equation}
		\label{Krn} 
		K = \frac{12r_0 ^2}{r^6} \left[1-\frac{4r_Q ^2}{r r_0} + \frac{14r_Q ^4}{3r_0 ^2 r^2}\right]~.
	\end{equation}
	Below, a necessary condition for the string scale to determine the magnitude of $K$ is written.
	Defining for $r\neq0$
	\begin{equation}
		z\equiv \frac{r_Q ^2}{r r_0}~,
	\end{equation}
	allows one to rewrite Eq.~(\ref{Krn}) as
	\begin{equation}
		K = \frac{56 r_0 ^8 }{r_Q ^{12}} z^6 \Big(z^2 -\frac{6}{7}z + \frac{3}{14}\Big)~.
	\end{equation} 
	This is a monotonically increasing function of $z$, implying that the curvature grows toward the interior. The quadratic function $z^2 -\frac{6}{7}z + \frac{3}{14}$ receives a global minimum at $z=\frac{3}{7}$ where it evaluates to $\frac{3}{98}$. It follows that
	\begin{equation}
		 K \geq \frac{12 r_0^2}{7r^6}~,
	\end{equation}
	implying that a necessary condition for stringy tidal forces is $r\lesssim r_*$. The case	 $\ell_s \ll r_-$ is considered to make the statement below non-trivial.  This implies that $\sqrt{\ell_s  r_0}\ll r_Q$.\\
	  An existence proof of string-sized tidal forces in a continuous family of charged black hole solutions is 
	\begin{equation}
		r \approx r_* ~,~ \sqrt{\ell_s r_0}\ll r_{Q} \ll \ell_s ^{\frac{1}{3}} r_0 ^{\frac{2}{3}}~ \Rightarrow K \approx \frac{1}{(\alpha')^2}~.
	\end{equation}
		 A comment is that the $\frac{ r_Q ^4}{r^8}$ behavior of $K$ in Eq.~(\ref{Krn}) for $r\ll \frac{r_Q ^2}{r_0}$ implies strong curvature at $r\sim \sqrt{r_Q \ell_s}$, however one can check that this lies behind the inner horizon. In GR, one expects perturbations would render the region behind the Cauchy horizon inaccessible \cite{Penrose} and generate a weak null singularity replacing the inner horizon, which can be interpreted as saying that the scale $\sqrt{r_Q \ell_s}$ is uninteresting. A numerical plot comparing the magnitude of $r_*$ and $r_-$ as a function of $\frac{r_Q}{r_0}$ with the choice $\frac{r_0}{\sqrt{\alpha'}}=100$ is presented.
		 \begin{figure}[h]
		 	\centering
		 	\includegraphics[scale=0.5]{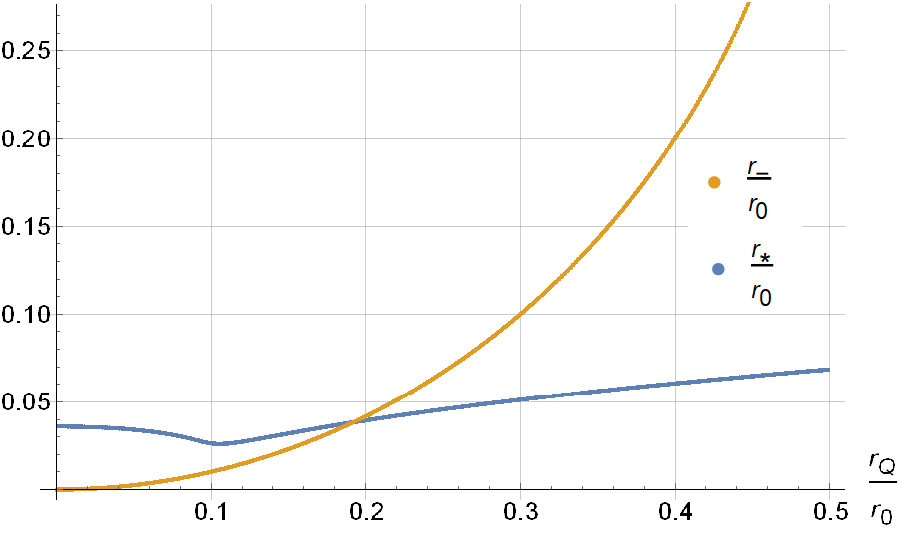}
		 	\caption{The orange curve is  $\frac{r_-}{r_0}$ as a function of the ratio $\frac{r_Q}{r_0}$. The blue curve is the greartest root of $K-\frac{1}{(\alpha')^2}$, denoted by $\frac{r_*}{r_0}$, as a function of $\frac{r_Q}{r_0}$, with the choice $\frac{r_0}{\sqrt{\alpha'}}=100$. When $\frac{r_Q}{r_0}\lesssim0.194$, $r_-<r_*$, which is the interesting range where effective field theory for infalling probes breaks down prior to reaching the inner horizon.  }
		 \end{figure}

	The conclusion of this subsection is that there is a range of charged black holes with interiors
	exhibiting stringy tidal forces outside (would-be) inner horizons greater than the string scale, yet several orders of magnitude smaller than the outer horizon.
		
	\subsection{D0-Brane Black Hole}
	\label{subs:D0}
	In this subsection, scales of strong curvature are written for the black hole dual to the thermal state of the large-$N$ BFSS gauged matrix model \cite{BFSS}. First, some basic properties of the string theory bulk are reviewed. Taking a decoupling limit of the black hole solution that carries $N$ D0-brane quantized charge \cite{Horowitz91} and turning on finite temperature lead to the following line element, dilaton, and RR one-form potential \cite{Itzhaki98},\cite{Xi13}
	\begin{equation}
		\label{TypeIIAD0}
		ds^2= -\frac{1}{\sqrt{f}(r)}A(r) dt^2+\frac{\sqrt{f(r)}dr^2}{A(r)}+ \sqrt{f(r)}r^2 d\Omega_8 ^2~,~ e^{\Phi} =  f(r) ^{\frac{3}{4}}~, 
	\end{equation}
	\begin{equation}
		C^{(1)} = -\frac{1+A(r)}{2f(r)}dt~,
	\end{equation}
	\begin{equation}
		f(r)\equiv \frac{c_0 g_s N \ell_s ^7}{r^7}~,~ c_0 = 60\pi^3~,~ A(r) \equiv 1-\frac{r_0 ^7}{r^7}~.
	\end{equation}
	In these equations, $r_0$ is the horizon's position, and $g_s$ is a coupling parameter. The 11th-dimensional Planck length is defined as
	\begin{equation}
		\ell_P \equiv (2\pi g_s)^{\frac{1}{3}}\ell_s~.
	\end{equation}
	The Hawking temperature of the black hole is 
	\begin{equation}
		T = \frac{7r_0 ^{\frac{5}{2}}}{4\pi\sqrt{c_0 g_s N} \ell_s ^{\frac{7}{2}} } ~.
	\end{equation} 
	The usual story told about the zero-temperature Type IIA bulk is that it is weakly coupled and weakly curved in an intermediate region of radii $r_{min}\ll r\ll r_{max}$, where
	\begin{equation}
		r_{min} = \frac{1}{(g_s N)^{\frac{4}{21}}} N^{\frac{1}{3}} \ell_P ~,~ r_{max} =  N^{\frac{1}{3}} \ell_P ~.
	\end{equation}
	When $g_s N\gg1$, the following hierarchy applies $r_{min} \propto (g_s N)^{\frac{1}{7}} \ell_s \gg \ell_s$.
 If a D0-brane propagates on the strong curvature region $r>r_{max}$, one can use the weakly coupled matrix model to describe its position as a function of time. In contrast, the brane enters strong coupling for $r\leq r_{min}$, and one should start uplifting to 11D supergravity.\\ However, at finite temperatures, the same story would be misleading for the following reason. The Ricci scalar, Weyl tensor squared, and Kretschmann scalar are given by
	\begin{equation}
		\label{RicciD0}
		R = \frac{49 \sqrt{\frac{3}{10}} (r^7 + 3 r_0 ^7)}{4 \pi   \sqrt{ N \ell_P ^3} \ell_s ^2 r^{\frac{11}{2}}}~,
	\end{equation}
	\begin{equation}
		C_{\alpha \beta \gamma \delta} C^{\alpha \beta \gamma \delta} = 
		\frac{7 \Big(7 r^7 - 6 r_0^7\Big)^2}{120\pi^2 N \ell_P ^3 \ell_s ^4   r^{11}}~,
	\end{equation}
	\begin{equation}
		K =\frac{7 \Big(175 r^{14} - 126 r^7 r_0^7 + 639 r_0^{14}\Big)}{240 \pi^2  N \ell_P ^3 \ell_s ^4   r^{11}}~.
	\end{equation}
	It is shown that for a sufficiently large $\frac{r_0}{\ell_P}$, an infaller first meets stringy tidal forces at a particular scale $r=r_* ^{(10D)}$, and then, strong coupling too. Consider the case where the horizon admits weak curvature, $r_0 \ll N^{\frac{1}{3}}\ell_P $; otherwise, the stringy tidal forces occur at or outside the horizon. What is the black hole interior sphere when tidal forces are stringy?  A simplification occurs if one can neglect the $r^7,r^{14}$ terms in the numerators of $R,C^2,K$ - which is justified shortly. 
	One thus obtains 
	\begin{equation}
		\label{U*}
		R \sim \frac{1}{\ell_s ^2}~\text{or}~C^2,K \sim  \frac{1}{\ell_s ^4} \Rightarrow r_* ^{(10D)} \sim r_0 \left(\frac{r_0}{  N^{\frac{1}{3}}\ell_P}\right)^{\frac{3}{11}}~.
	\end{equation}
	The assumption of weakly curved horizon, $r_0 \ll N^{\frac{1}{3}}\ell_P $, implies that $r_* ^{(10D)}\ll r_0$ which means that the approximation of neglecting the $r^7,r^{14}$ terms in the numerators of $R,C^2,K$ is controlled.
	 Moreover, a calculation of the $(\alpha')^3 R^4$ correction in the Type IIA effective action for the bulk solution in Eq.~(\ref{TypeIIAD0})  gives rise to
	\begin{align}
		&\frac{e^{2\Phi}}{\sqrt{-G}}L_{R^4} ^{II}=\frac{\zeta(3) (\alpha')^3}{32 \kappa_0 ^2}  \Big( 2 R_{\mu \nu \rho \sigma} R_{\alpha ~~~\beta} ^{~~\nu \rho} + R_{\mu \sigma \nu \rho } R_{\alpha \beta} ^{~~~\nu \rho} \Big)R^{\mu \gamma \delta \alpha} R^{\beta~~~~ \sigma} _{~~\gamma \delta}~\nonumber\\
		  &= \frac{7\zeta(3)}{32\kappa_0^2}  \frac{133427 r^{28} + 360836 r^{21} r_0^7 - 310366 r^{14} r_0^{14} + 
		 159012 r^7 r_0^{21} + 
		 3262851 r_0^{28}}{409600\pi^4  \ell_P^6 (\alpha') N^2 r^{22}}~.
	\end{align}
	The Einstein-Hilbert Lagrangian density in Eq.~(\ref{RicciD0})	is comparable to $\frac{e^{2\Phi}}{\sqrt{-G}}L_{R^4} ^{II}$ at $r\approx 0.7r_*$.
	 This implies that effective field theory is reliable provided $r\gg r_*$.
	
	Could one encounter $r=r_* ^{(10D)} \gg  r_{min}$, namely strong curvature at weak string coupling? The answer is positive for
	\begin{equation}
		\label{r0Bound}
		\frac{1}{(g_s N)^{\frac{22}{147}}} N^{\frac{1}{3}}\ell_P \ll r_0~.
	\end{equation} 
	A complementary regime is solutions with $\frac{1}{(g_s N)^{\frac{4}{21}}} N^{\frac{1}{3}}\ell_P\ll r_0 < \frac{1}{(g_s N)^{\frac{22}{147}}} N^{\frac{1}{3}}\ell_P$ where the standard assumption made in the literature of strong coupling first holds true.~\footnote{ Another constraint that one could take into account is that the Hawking temperature of the D0-brane black hole is below the Hagedorn temperature of flat spacetime, $T< T_H ^{II} =\frac{1}{2\pi \sqrt{2\alpha'}}$, to avoid divergence of the one-loop partition function computed at the weakly curved and weakly coupled region. This sets an upper bound on the horizon size $r_0 \leq \frac{N^{\frac{1}{3}} \ell_P}{(g_s N)^{\frac{2}{15}}}$; this upper bound is greater than the scale on the L.H.S of Eq.~(\ref{r0Bound}) and smaller than $N^{\frac{1}{3}}\ell_P $.}  
	\begin{figure}[h!]
		\begin{center}
			\includegraphics[scale=0.75]{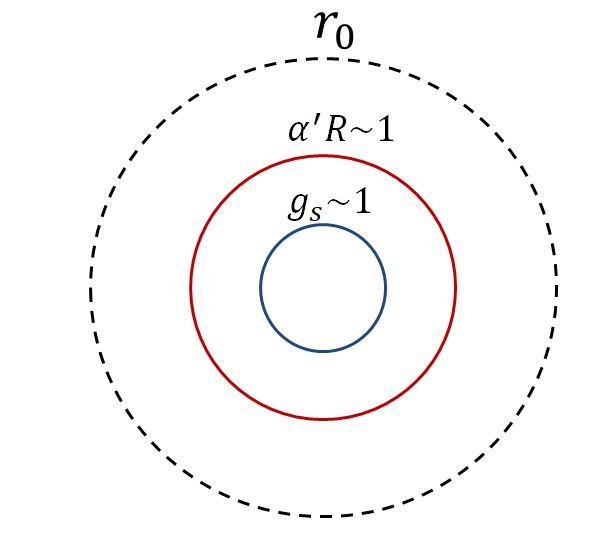}
			\caption{In solutions with $	\frac{1}{(g_s N)^{\frac{22}{147}}} N^{\frac{1}{3}}\ell_P \ll r_0$,  low-energy infallers that do not carry momentum along the 11th-dimensional circle become strings at the red region before meeting the order one string coupling region.}
					\label{fig:D0BH}
		\end{center}
	\end{figure}
	\\
	Next, consider an infaller that carries nonzero momentum along the 11th-dimensional circle. 
	The line element reads \cite{Xi13}
	\begin{equation}
		\label{11up} 
		ds_{11} ^2 = f(r) dx_{11} ^2 - (1+A(r))dx_{11} dt + \frac{(1-A(r))^2}{4f(r)}dt^2 + \frac{dr^2}{A(r)} + r^2 d\Omega_8 ^2~.
	\end{equation}
	Quantum gravitational effects are important in the black hole interior, provided the 11th-dimensional curvature invariants are Planckian. The Ricci tensor of Eq.~(\ref{11up}) vanishes; however, the Kretschmann scalar and the Weyl squared do not
	\begin{equation}
		\label{Kretch11} 
		K= C^2 = \frac{4032 r_0 ^{14}}{r^{18}}~.
	\end{equation}
	\footnote{I thank Jorge Santos for emphasizing the importance of working in 11D and sharing an efficient code that computes Eq.~(\ref{Kretch11}). } Therefore, a cluster of nearby and infalling D0-branes feels Planckian tidal forces at
	\begin{equation}
		r_* ^{(11D)} \approx 1.58  r_0 ^{\frac{7}{9}} \ell_P ^{\frac{2}{9}}~.
	\end{equation}  
	This is precisely the scale at which an infaller in a 10D Schwarzschild-Tangherlini black hole interior encounters Planckian curvature. Assuming that the 11D solution is stable against the Gregory–Laflamme instability requires $r_0 \gg N^{\frac{1}{9}} \ell_P$ \cite{Polchinski99}, which implies $r_* ^{(11D)}\gg N^{\frac{7}{81}} \ell_P$. A stronger requirement, that $r_0 \gg  N^{\frac{9}{49}}\ell_P $, implies that the scale of stringy tidal forces precedes the scale of Planckian tidal forces $r_{*} ^{(11D)}<r_* ^{(10D)}$; otherwise, the hierarchy of scales is reversed and a cluster of infalling gravitons would experience tidal disruption first at $r_* ^{(11D)}$.
	
	If renormalization group flow in the matrix quantum mechanics system can be thought of as the radial bulk coordinate, and strong curvature in the bulk amounts to weak coupling in BFSS, then the Yang-Mills coupling constant would become weak below the energy scale corresponding to $r_* ^{(10D)}$. One would retain a perturbative treatment in terms of classical matrices, which describe the region between the red and blue circles in Fig.~\ref{fig:D0BH}~. Open string excitations that stretch between the D-particles would be heavy, and a one-loop effective potential would determine the trajectories of dynamical D0-branes in that region. This interpretation implies that a portion of the D0-brane black hole interior is characterized in the dual system by a Coulomb phase of D0-branes distributed sparsely in space. 
	The similarity of the physics of the branes very far away from the black hole region and within the intermediate interior region suggests that the thermal state in BFSS displays a non-trivial degree of entanglement between high-energy and low-energy D-branes. 
	
	\subsection{Black p-Branes}
	\label{subs:Dp}
	This subsection generalizes the above results for the Horowitz-Strominger black p-brane solutions \cite{Horowitz91} in the decoupling limits at finite temperature, which were written by \cite{Itzhaki98}. Reference \cite{Horowitz1997} showed that the solutions  admit divergent Riemann tensor components in frames describing massive, radially infalling modes that approach the horizon. The papers \cite{Emil98a},\cite{Emil98b} wrote conditions of the validity of effective field theory at the horizon scale and characterized the phase diagrams of the dual super-Yang Mills theories. On the other hand, scales of breakdown of effective field theory for any frame in the interiors of these solutions are written in this subsection.  Here is a brief review of these backgrounds.
	The solutions in question are spherically symmetric with a $\mathbb{T}^p$ compactification manifold, with quantized RR charge $N$, and below, the set $p=\{0,...,6\}$ is considered. The cases $p\neq3$ are analyzed first; the case $p=3$ is deferred to the end of the subsection.
	The exponential of the dilaton is given by
	\begin{equation}
		e^{\Phi} = (2\pi) ^{2-p} \frac{\lambda}{N} \left( \frac{\lambda d_p}{r^{7-p}}\right)^{\frac{3-p}{4}}~,
	\end{equation}
	with
	\begin{equation}
		\lambda =  (2\pi) ^{p-2} g_s N \ell_s ^{p-3}~,~
		d_p \equiv 2^{7-2p} \pi ^{\frac{9-3p}{2}} \Gamma\left(\frac{7-p}{2}\right)~.
	\end{equation}
	 The line element of the solutions in the string frame is given by  
	\begin{equation}
		\frac{1}{\alpha'}ds^2 = -f(r) dt^2 + \frac{dr^2}{f(r)} + g(r) d\Omega_{8-p} ^2 + h(r) ds_{\mathbb{T}^p} ^2~, 
	\end{equation}
	where
	\begin{equation}
		f(r) \equiv \frac{r^{\frac{7-p}{2}}}{\sqrt{\lambda d_p}} \left(1-\frac{r_0^{7-p}}{r^{7-p}}\right)~,~
		g(r) \equiv \sqrt{\lambda d_p}  r^{\frac{p-3}{2}}~,~
		h(r) \equiv \frac{r^{\frac{7-p}{2}}}{\sqrt{\lambda d_p}}~.
	\end{equation}
      The Hawking temperature is given by
      \begin{equation}
      	T = \frac{(7-p) r_0 ^{\frac{5-p}{2}}}{4\pi \sqrt{d_p \lambda}}~.
      \end{equation}
      Scales corresponding to strong coupling, stringy torus size, and strong curvature are written below, as they are important in dissecting when effective field theory breaks down.
    
     	The dilaton increases monotonically with $r$ for $p>3$, making string theory strongly coupled far away; by contrast, weaker coupling is encountered toward large values of $r$ when $p<3$. A significant scale is where the string coupling is of order one:
     \begin{equation}
     	\label{Ugs1}
     	r\sim r_{g_s\sim 1}(p) \equiv\left[d_p (2\pi)^{\frac{4(2-p)}{3-p}}\right]^{\frac{1}{7-p}}\frac{\lambda^{\frac{1}{3-p}}}{N^{\frac{4}{(3-p)(7-p)}}}~.
     \end{equation}
     For $p\in \{1,...,6\}$,  let us denote the asymptotic volume of the torus by $V$ and assume that all these compact dimensions are of equal length $V^{\frac{1}{p}}$.
     Since $h(r)$ increases with $r$, the proper torus size decreases toward the space interior. The sizes of the torus cycles are comparable to the string length at
     \begin{equation}
     	r\sim r_{small~\mathbb{T}^p} (p)\equiv  V^{-\frac{4}{p(7-p)}}\Big( d_p\lambda\Big)^{\frac{1}{7-p}}~.
     \end{equation}
     For a sufficiently large $V$, this scale is in the black brane region: $r_{small~\mathbb{T}^p} (p)<r_0$. The geometry's sphere factor $S^{8-p}$ expands far away if $p>3$ and shrinks to zero asymptotically for $p<3$, where strong curvature awaits scattering states.  
     The Ricci scalar of the geometry is 
     \begin{align}
     	&R = -\frac{(p-5)(p-8) f (g')^2}{4\alpha' g^2} +\frac{p(p-8) f g' h'}{2\alpha' gh} + \frac{(p-7)(p-8)}{\alpha' g}-\frac{f''}{\alpha'} -\frac{p f h''}{\alpha' h} \nonumber\\
     	& ~~~~+\frac{(p-8)f' g'}{\alpha' g}+\frac{(p-8) f g''}{\alpha' g}-\frac{p f'h'}{\alpha' h} - \frac{p(p-3)f (h')^2}{4\alpha' h^2} -\frac{p(p-3) f (h')^2}{4\alpha' h^2} \nonumber\\
     	& ~~~~~~=\frac{(p-7)^2 (3-p)}{4\alpha' \sqrt{d_p \lambda}} r^{-\frac{11+p}{2}} \left((p+1)r^7+(3-p) r^p  r_0 ^{7-p}\right)~.
     \end{align}
      As a check, one recovers Eq.~(\ref{RicciD0}) for $p=0$. 
      The horizon is weakly curved for
      \begin{equation}
      	r_0 \ll \lambda^{\frac{1}{3-p}}~ (p<3) ~\text{or}~ r_0 \gg  \lambda^{\frac{1}{p-3}} ~~ (p>3)~.
      \end{equation}
      Here is the most important sentence in this subsection: The black p-brane interior is strongly curved when
      \begin{equation}
      	r\sim r_{strong}(p) \equiv  \left[ \frac{(p-7)(3-p)}{2} \right] ^{\frac{4}{11-p}}r_0 \left(\frac{r_0}{\lambda^{\frac{1}{3-p}}}\right)^{\frac{3-p}{11-p}}~.
      \end{equation}
      It is shown that the onset of the unreliability of effective field theory is at $r_{strong}(p)$. There are two possibilities. The first possibility is that
      \begin{equation}
      	\label{1stPoss}
      	r_{small~\mathbb{T}^p}(p)<r_{strong}(p) ~.
      \end{equation} 
       If also $p<3$, Eq.~(\ref{Ugs1}) permits one to take $N$ large enough such that $r_{g_s \sim 1}(p)\ll 	r_{strong}(p)$. Then effective field theory is valid for $r\gg r_{strong}(p)$, at weak string coupling and large torus cycles. If, on the other hand, $p>3$, then at large $N$, the string coupling becomes of order one far away from the black brane horizon, but the scale of ``naive strong curvature'' obtained at zero temperature, $\lambda^{\frac{1}{3-p}}$, is much smaller than $r_{strong}(p)$ when the horizon is weakly curved. One thus arrives at the same conclusion as in the $p<3$ case that effective field theory is reliable for $r\gg r_{strong}(p)$.  \\ The second possibility is that  
      \begin{equation}
      		r_{small~\mathbb{T}^p}(p)>r_{strong}(p)~.
      \end{equation}
      In this case, one could conclude that effective field theory breaks down at $r_{small~\mathbb{T}^p} (p)$. Alternatively, at the scale $r_{small~\mathbb{T}^p} (p)$, one can connect a bulk obtained by T-dualizing along all torus directions, making the volume of the $\mathbb{T}^p$ larger than the string scale to the power $p$. This also modifies the dilaton to be stronger as one dives into the geometry
      \begin{equation}
      	e^{\Phi'} = \frac{e^{\Phi}}{Vh(r)^{\frac{p}{2}}}=\frac{(2\pi)^{2-p}}{N} \lambda^{\frac{7}{4}} \frac{d_p ^{\frac{3}{4}}}{ Vr^{\frac{3(7-p)}{4}}}~. 
      \end{equation}  
      It follows that effective field theory based on the T-dual frame becomes strongly coupled at
      \begin{equation}
      	r\sim r_{g_s\sim 1} '(p) \equiv \left[(2\pi)^{\frac{4(2-p)}{3}}d_p\frac{\lambda^{\frac{7}{3}}}{(NV) ^{\frac{4}{3}}}\right]^{\frac{1}{7-p}} ~.
      \end{equation}
      Computing the Ricci scalar of the T-dualized geometry, one obtains
      \begin{equation}
      	R = \frac{3(p-7)^2 r^{-\frac{11+p}{2}}\Big( r^7 + 3 r_0 ^{7-p} r^p\Big)}{4\alpha' \sqrt{d_p \lambda}}~.
      \end{equation}
      Therefore, for a weakly curved horizon, $R\sim \frac{1}{\alpha'}$ at $r\sim r_{strong}' (p)=\left[ \frac{(p-7)3}{2} \right] ^{\frac{4}{11-p}}r_0 \left(\frac{r_0}{\lambda^{\frac{1}{3-p}}}\right)^{\frac{3-p}{11-p}}$. For $p>3$, $r_{strong}'(p)$ is greater than the scale of strong curvature at zero temperature.
      Taking $N$ to be significantly larger than any other dimensionless parameter in the theory implies that string theory remains weakly coupled in a portion of the black brane interior, both in the original and T-dual frames. Consequently, the black p-brane interior region does not admit a good effective field theory description when   
      $r\sim r_0 \left(\frac{r_0}{\lambda^{\frac{1}{3-p}}}\right)^{\frac{3-p}{11-p}}$.\\
      Next, let us consider the $p=3$ case relevant to the correspondence between Type IIB superstring theory on asymptotically $AdS_5 \times S^5$ and four-dimensional $SU(N)$ superconformal Yang-Mills gauge theory on $\mathbb{T}^3 \times R$.  The total Ricci scalar of the bulk solution vanishes, but the Kretschmann scalar is non-trivial:
      \begin{equation}
      	K(p=3) = \frac{40 r^8+36 r_0 ^{8}}{ r^8 (\alpha')^2 \lambda}~.
      \end{equation}
      The horizon is weakly curved for $\lambda \gg 1$, which is assumed. The torus becomes stringy at $r\sim V^{-\frac{1}{3}}\lambda^{\frac{1}{4}}$.
      The black threebrane interior is strongly curved at
      \begin{equation}
      	r_{strong} (p=3) \approx \frac{1.56 r_0}{\lambda^{\frac{1}{8}}}~.
      \end{equation}
      This occurs for weak string coupling ($\lambda \ll N$). If the torus arrives at a string-sized volume at a scale larger than $r_{strong}(p=3)$, one can T-dualize to reach a frame where the torus is large, and the conclusion is that the lower limit of validity of effective field theory is  $r_{strong}(p=3)$.\\ 
      Two comments are in order. First, the scale of Planckian curvature occurs at $g_s^{\frac{1}{11-p}}r_{strong} (p)$ which is well-inside $r_{strong} (p)$ for weak string coupling. Second, the phenomenon of Gregory-Laflamme instability takes place in a regime of parameter space when the entropy satisfies $S<V^{\frac{9}{2}}N^{\frac{1}{2}}$ \cite{Emil98b} when the size of the torus dimensions at the horizon is larger than the object itself.  In that case, the system transitions to the unsmeared black hole solution with horizon localized on the $\mathbb{T}^p$, or the D0-brane black hole that was the topic of the previous subsection - where it was explained that for large $\frac{r_0}{\ell_p}$, strong curvature happens in the interior prior to losing the $g_s$-expansion.

      A paragraph about the paper \cite{Horowitz1997} is written. It showed the existence of Planckian tidal forces on infalling probes near the horizons of extremal and near-extremal black p-branes solutions of \cite{Horowitz91}. While the authors did not take the decoupling limit, this effect persists in this limit: $R_{riri}\propto \frac{1}{r-r_0}$ near $r=r_0$ with $i$ being an angular direction or a torus direction. A massless test string and a massive string in an infalling frame feel these tidal forces near the black p-brane horizon. The results of the present subsection are different than \cite{Horowitz1997} in that they are a)  relevant for the interior rather than the exterior, b) specifying when the $\alpha'$ expansion breaks down, whereas the paper above explained that $\alpha'$ corrections are unimportant in the exterior. \\
      A summary of the subsection results appears in Table \ref{Table1} relevant to the possibility (\ref{1stPoss}).
    \begin{table}[h!]
    	\centering
\begin{tabular}{|p{2cm}|p{2cm}|p{2cm}|p{2cm}|}
	\hline
	$p$& Naive\linebreak Minimal $r$ & Minimal $r$& Maximal $r$\\
	\hline
	Black zerobrane   & $\frac{\lambda^{\frac{1}{3}}}{N^{\frac{4}{21}}}$    & $r_0 \left(\frac{r_0}{\lambda^{\frac{1}{3}}}\right)^{\frac{3}{11}}$  &   $\lambda^{\frac{1}{3}}$\\
	\hline
	 Black onebrane& $\frac{\lambda ^{\frac{1}{2}}}{N^{\frac{1}{3}}}$    & $r_0 \left(\frac{r_0}{\lambda^{\frac{1}{2}}}\right)^{\frac{1}{5}}$   &$\lambda^{\frac{1}{2}}$\\
	\hline
	Black twobrane & $\frac{\lambda}{N^{\frac{4}{5}}}$ & $r_0 \left(\frac{r_0}{\lambda}\right)^{\frac{1}{9}}$&  $\lambda$\\
	\hline
	Black threebrane   &$0$ & $ \frac{r_0}{\lambda^{\frac{1}{8}}}$&  $\infty$\\
	\hline
	Black fourbrane &   $\frac{1}{\lambda}$  & $\frac{r_0}{ \Big(\lambda r_0\Big)^{\frac{1}{7}}}$&$\frac{N^{\frac{4}{3}}}{\lambda} $\\
	\hline
	Black fivebrane & $\frac{1}{\lambda ^{\frac{1}{2}}}$  & $\frac{r_0}{ \Big(\lambda ^{\frac{1}{2}} r_0\Big)^{\frac{1}{3}}}$   &$\frac{N}{\lambda^{\frac{1}{2}}}$\\
	\hline
	Black sixbrane& $\frac{1}{\lambda^{\frac{1}{3}}}$  & $\frac{r_0}{ \Big(\lambda^{\frac{1}{3}} r_0\Big)^{\frac{3}{5}}}$&$\frac{N^{\frac{4}{3}}}{\lambda^{\frac{1}{3}}}$\\
	\hline
\end{tabular}
\caption{The leftmost column corresponds to the value of $p$ of the $p$-brane solution, to its right is the scale obtained when the coupling is strong for $p<3$, or the curvature at zero temperature is strong for $p>3$. The rightmost column presents the scale of strong curvature for $p<3$ and strong coupling for $p>3$. The column to its left is the list of scales of strong curvature in black p-brane interiors. ~\label{Table1}}
\end{table}
The conclusion is that effective field theory breaks down in black brane interiors where the curvature is strong in string units, the coupling is weak, and the torus size is large in string units. \newpage

	\section{Summary and Final Comments}
	\label{subs:Summ}
	In this paper, I showed that there are macroscopic scales in which target space effective field theories in black hole interior backgrounds break down in tandem with string-sized tidal forces, where the worldsheet theory on the sphere becomes strongly coupled. Alternatively, if there is a compact dimension transverse to the black hole dimensions, which is somewhat larger than the string scale, then the effective field theory would break down at a macroscopic scale $r_{*int}$ set by that extra dimension. A third scale corresponds to when tidal forces in the black hole interior are comparable to the Planck scale. For the Schwarzschild interior, one obtains a scale much greater than the Planck length, $r_{*P}=(\ell_P ^2 r_0)^{\frac{1}{3}}\gg \ell_P $. If string theory is irrelevant to nature and there are no extra dimensions, then quantum gravity is important already at $r_{*P}$.  If string theory is relevant to nature, $r_{*P}$ is the earliest among $r_*,r_{*int}$ provided the string coupling is much greater than one. One can interpret these scales as times of ``drama'' in the black hole interior.
	
	This physical picture differs from a GR prediction for an imaginative, resilient unicellular organism that falls to the black hole region, experiencing enormous acceleration and tidal forces until it dies at the singularity. GR breaks down because the structure of the higher-derivative contributions to the target space effective action includes powers of the Riemann tensor, and these become stringy in the interior at $r_*$. In the exterior, the scale of breakdown is $\sqrt{\alpha'}$, as exemplified by Type II corrections to thermodynamic quantities of Schwarzschild, which include a term proportional to $\sim \frac{(\alpha')^3}{r_0 ^6}$ \cite{Myers1987},\cite{Yiming}.  
	
	The physical picture of Itzhaki et al~ \cite{Itzhaki98} for zero-temperature black p-branes, which includes a minimal radial coordinate in target space where effective field theory is lost, has been extended in the present paper for finite temperature. Adopting the \cite{Itzhaki98} picture with the minimal, though macroscopic, bulk radius for the Schwarztschild, Kerr and Reissner-N\"ordstrom black holes is also useful.
	
	One implication of this study is that infalling low-energy probes become string loops when entering the scale $r_*$. This is also valid for massless Hawking modes that start their lives in the black hole interior and travel toward the central region. The strings stretch and become trapped by gravity. If the tidal forces first correspond to the size of the extra dimensions, then the probe is converted into winding or momentum modes about these dimensions. Therefore, the black hole interior can be viewed as a particle-to-string accelerator!
	
	In the context of the gauge/string correspondence, if it applies to the black hole interior as well,  a quantum field theoretic interpretation of the experience of the infalling bulk probe is that a thermal state is added with a quench corresponding to a single-trace operator with a nonzero expectation value, which loses its energy or scrambles to multi-trace operators. Since bulk regions of strong curvature are associated with an RG flow to weak coupling on the gauge theory side, energy is partitioned at a much slower rate among the degrees of freedom of the gauge theory system relative to the rate at strong coupling. Understanding this further and other aspects of the interpretation would be interesting.
	
	Arguments that effective field theory behind the horizon describes ``low-complexity'' operators to a good approximation \cite{Non},\cite{Strings2024} should be regulated given the results of the present paper because these claims did not incorporate the increase of the complexity of probes in the interior at macroscopic scales like $r_{*P},r_{*}$. Similarly, the ER=EPR idea \cite{Maldacena13} should incorporate a conclusion of this paper that the semiclassical approximation does not hold at $r=r_*$.  
		
	A few speculations about black holes formed by collapsing shells are written. The rules of string theory imply that if the shell reaches $r_*$, it transitions to strings. A subset of the strings is expected to collide with each other, trading energy between themselves and emitting short strings that are likely to return toward the center.  
	
	The black hole singularity could be resolved in the process of the gravitational collapse since highly-excited strings display soft (fixed-angle) scattering \cite{Gross87}; a classical spacelike singularity, on the other hand, can be characterized as an obstruction to scattering - it absorbs completely infalling matter. 
	
	Reference \cite{Mathur:2008kg} presented the idea that the singularity is avoided because when the shell reaches the horizon scale, it has a tunneling probability of $e^{-S}$ (where $S$ is the black hole entropy) to make a transition to a fuzzball. There are $e^{+S}$ such states, imputing an order one probability it transitions to a singularity-free bound state of string theory. Each state would be stable against collapse because of fluxes that counterbalance gravity or quantum effects that lead to a regular wavefunction for the bound state. However, how to perform rigorous calculations for uncharged, non-supersymmetric systems that would confirm or rule out these ideas is unknown.  
	
	If the ideas in \cite{Mathur:2008kg} are irrelevant to black hole physics, but string theory still applies, then after a light crossing time, the state of the initial collapsing shell is unknown since the string coupling can become strong in the interior (which, e.g., happens in the 2D black hole of \cite{Witten91}), where light D- and NS-branes can be created. I expect that quantum effects for these light branes would eventually resolve the singularity similarly to the resolution of the singularity of the planetary model of the atom, where the uncertainty principle and the virial theorem imply a finite support for the wavefunction of the atom, one that is distinct from a delta-function in position space.

	\section*{Acknowledgements}
	I thank Amr Ahmadain, David Benisty, Alexander Frenkel, Robie Hennigar, \linebreak Emil Martinec, Harvey Reall,  Jorge Santos,  Watse Sybesma, Xi Tong and Gabriele Veneziano for discussions. I thank Sunny Itzhaki and Gary Horowitz for their critical comments on previous versions of the manuscript. YZ is supported by the Blavatnik fellowship.
 
	\appendix
	\section{Low-Dimensional Black Holes}
	
	\subsection*{3D}
	The BTZ black hole can be written as an exact two-dimensional CFT on the sphere worldsheet given by the WZW coset model $SL(2,R)_k / Z_2$ \cite{Horowitz93},\cite{Natsuume:1996ij} with 
	\begin{equation}
		k =\frac{L^2}{\alpha'} ~,~
	\end{equation}
	\begin{equation}
		\label{BTZ}
		ds^2 = \left(M - \frac{r^2}{L^2}\right)dt^2 - J dt d\phi  +r^2 d\phi^2 +\frac{dr^2}{\frac{r^2}{L^2}-M +\frac{J^2}{4r^2}}~,
	\end{equation}
	\begin{equation}
		B^{(2)} = \frac{r^2}{L}  d\phi \wedge dt~.
	\end{equation}
	The dilaton is constant. A higher-dimensional example where such a geometry arises is the D1-D5 extremal, rotating black hole solution to the leading order supergravity equations, which contains a BTZ factor.
	The geometry in Eq.~(\ref{BTZ}) has a constant Ricci scalar everywhere $R= -\frac{6}{L^2}$ and a constant Kretschmann as well $K = \frac{12}{L^4}$; thus, this system does not display stringy tidal forces unless $L\leq \ell_s$. In the latter case, black holes are non-normalizable -  outside the spectrum \cite{Giveon:2005mi} and tidal forces are everywhere string-sized. 
	
	Another example of curvature constancy is the two-dimensional Jackiw–Teitelboim gravity black hole, where the Ricci scalar is proportional to the cosmological constant and the Kretschmann scalar to its square. 

	\subsection*{2D} 
	A two-dimensional black hole with asymptotically linear dilaton admits an exact CFT description \cite{Bars:1990rb},\cite{Witten91},\cite{Dijkgraaf:1991ba}. The supersymmetric version of the coset model is considered. 
	Using expressions in \cite{Giveon:2005mi}, the metric and dilaton are given by
	\begin{equation}
		ds^2 = -f(r) dt^2 + \frac{k\alpha' dr^2}{2r^2 f(r)}~,~ \Phi (r)= - \frac{1}{2} \log\left(\frac{\sqrt{k}r}{\sqrt{2\alpha'}}\right)~,
	\end{equation}
	\begin{equation}
		f(r)\equiv 1-\frac{2M \alpha'}{r}~,~ M = \frac{1}{\sqrt{2k\alpha'}g_s ^2}~. 
	\end{equation}
	The horizon is at $r=2M \alpha'$, the mass is $M$ and $k$ is the level of the  $SL(2,R)_k/U(1)$ gauged WZW model. The string coupling at the horizon is $g_s$. The Ricci and Kretschmann scalars are given by
	\begin{equation}
		R = \frac{4M}{kr}~,K= R^2~.
	\end{equation}
	The curvature becomes stringy at 
	\begin{equation}
		r_* = \frac{4M\alpha'}{k}= \frac{2^{\frac{3}{2}}\ell_s }{g_s ^2 k^{\frac{3}{2} }}~,
	\end{equation}
	which is parametrically greater than the string scale provided that $g_s \ll \frac{1}{k^{\frac{3}{4}}}$. In this case, the scale of order one string coupling, $r_{g_s\sim 1} = \ell_s \sqrt{\frac{2}{k}}$, is much smaller than $r_*$: $r_*\gg r_{g_s\sim 1}$.  However, unlike higher dimensional cases considered in the body of the paper, the target space does not receive $\alpha'$ corrections. It would be interesting to utilize the exact CFT description to find explicit expressions for the occupation numbers of string modes produced at $r=r_*$. The genus-zero CFT description is reliable when $r\gg r_{*P}\equiv g_s r_*$ where the curvature on infalling probes is sub-Planckian. The lower limit where one should view the CFT description on the sphere as reliable, $r_{*P}$, is much greater than $r_{g_s\sim1}$ and $\ell_P$ if $g_s \ll \frac{1}{k^{\frac{6}{5}}}$.


\begin{thebibliography}{99}

\bibitem{Hawking75}
S.~W.~Hawking,
``Particle Creation by Black Holes,''
Commun. Math. Phys. \textbf{43}, 199-220 (1975)
[erratum: Commun. Math. Phys. \textbf{46}, 206 (1976)]

\bibitem{Itzhaki96}
N.~Itzhaki,
``Is the black hole complementarity principle really necessary?,''
[arXiv:hep-th/9607028 [hep-th]].

\bibitem{Mathur09}
S.~D.~Mathur,
``The Information paradox: A Pedagogical introduction,''
Class. Quant. Grav. \textbf{26}, 224001 (2009)
[arXiv:0909.1038 [hep-th]].
\bibitem{Polchinski12}
A.~Almheiri, D.~Marolf, J.~Polchinski and J.~Sully,
``Black Holes: Complementarity or Firewalls?,''
JHEP \textbf{02}, 062 (2013)
[arXiv:1207.3123 [hep-th]].



\bibitem{Casher:1996ct}
A.~Casher, F.~Englert, N.~Itzhaki, S.~Massar and R.~Parentani,
``Black hole horizon fluctuations,''
Nucl. Phys. B \textbf{484}, 419-434 (1997)
[arXiv:hep-th/9606106 [hep-th]].

\bibitem{Sorkin:1996sr}
R.~D.~Sorkin,
``How wrinkled is the surface of a black hole?,''
[arXiv:gr-qc/9701056 [gr-qc]].

\bibitem{Marolf:2003bb}
D.~Marolf,
``On the quantum width of a black hole horizon,''
Springer Proc. Phys. \textbf{98}, 99-112 (2005)
[arXiv:hep-th/0312059 [hep-th]].

\bibitem{Bousso:2023kdj}
R.~Bousso and G.~Penington,
``Islands Far Outside the Horizon,''
[arXiv:2312.03078 [hep-th]].

\bibitem{Banks:2024imv}
T.~Banks, P.~Draper and M.~Karydas,
``Breakdown of field theory in near-horizon regions,''
JHEP \textbf{06}, 153 (2024)
[arXiv:2401.03572 [hep-th]].

\bibitem{Ong:2023lbr}
Y.~C.~Ong,
``A maximum force perspective on black hole thermodynamics, quantum pressure, and near-extremality,''
Eur. Phys. J. C \textbf{83}, no.11, 1068 (2023)
[arXiv:2309.04110 [gr-qc]].

\bibitem{Fuzzball}
I.~Bena, E.~J.~Martinec, S.~D.~Mathur and N.~P.~Warner,
``Fuzzballs and Microstate Geometries: Black-Hole Structure in String Theory,''
[arXiv:2204.13113 [hep-th]].


\bibitem{Maldacena13}
J.~Maldacena and L.~Susskind,
``Cool horizons for entangled black holes,''
Fortsch. Phys. \textbf{61}, 781-811 (2013)
[arXiv:1306.0533 [hep-th]].

\bibitem{Penington:2019kki}
G.~Penington, S.~H.~Shenker, D.~Stanford and Z.~Yang,
``Replica wormholes and the black hole interior,''
JHEP \textbf{03}, 205 (2022)
[arXiv:1911.11977 [hep-th]].

\bibitem{Almheiri:2019qdq}
A.~Almheiri, T.~Hartman, J.~Maldacena, E.~Shaghoulian and A.~Tajdini,
``Replica Wormholes and the Entropy of Hawking Radiation,''
JHEP \textbf{05}, 013 (2020)
[arXiv:1911.12333 [hep-th]].


\bibitem{Emil94}
E.~J.~Martinec,
``Space - like singularities and string theory,''
Class. Quant. Grav. \textbf{12}, 941-950 (1995)
[arXiv:hep-th/9412074 [hep-th]].

\bibitem{Horowitz90}
G.~T.~Horowitz and A.~R.~Steif,
``Strings in Strong Gravitational Fields,''
Phys. Rev. D \textbf{42}, 1950-1959 (1990)






\bibitem{Myers96}
N.~Kaloper, R.~C.~Myers and H.~Roussel,
``Wavy strings: Black or bright?,''
Phys. Rev. D \textbf{55}, 7625-7644 (1997)
[arXiv:hep-th/9612248 [hep-th]].
\bibitem{Ross97}
S.~F.~Ross,
``Singularities in wavy strings,''
JHEP \textbf{08}, 003 (1998)
[arXiv:hep-th/9710158 [hep-th]].
\bibitem{Maeda99}
K.~Maeda, T.~Torii and M.~Narita,
``String excitation inside generic black holes,''
Phys. Rev. D \textbf{61}, 024020 (2000)
[arXiv:gr-qc/9908007 [gr-qc]].


\bibitem{Silverstein12}
N.~Bao, X.~Dong, S.~Harrison and E.~Silverstein,
``The Benefits of Stress: Resolution of the Lifshitz Singularity,''
Phys. Rev. D \textbf{86}, 106008 (2012)
[arXiv:1207.0171 [hep-th]].

\bibitem{Arnold:2012qg}
P.~Arnold, P.~Szepietowski, D.~Vaman and G.~Wong,
``Tidal stretching of gravitons into classical strings: application to jet quenching with AdS/CFT,''
JHEP \textbf{02}, 130 (2013)
[arXiv:1212.3321 [hep-th]].

\bibitem{Silverstein14}
E.~Silverstein,
``Backdraft: String Creation in an Old Schwarzschild Black Hole,''
[arXiv:1402.1486 [hep-th]].


\bibitem{Emil20}
E.~J.~Martinec and N.~P.~Warner,
``The Harder They Fall, the Bigger They Become: Tidal Trapping of Strings by Microstate Geometries,''
JHEP \textbf{04}, 259 (2021)
[arXiv:2009.07847 [hep-th]].


\bibitem{Tyukov:2017uig}
A.~Tyukov, R.~Walker and N.~P.~Warner,
``Tidal Stresses and Energy Gaps in Microstate Geometries,''
JHEP \textbf{02}, 122 (2018)
[arXiv:1710.09006 [hep-th]].

\bibitem{Ceplak:2021kgl}
N.~Ceplak, S.~Hampton and Y.~Li,
``Toroidal tidal effects in microstate geometries,''
JHEP \textbf{03}, 021 (2022)
[arXiv:2106.03841 [hep-th]].

\bibitem{Guo:2021gqd}
B.~Guo and S.~Hampton,
``The dual of a tidal force in the D1D5 CFT,''
JHEP \textbf{07}, 149 (2023)
[arXiv:2108.00068 [hep-th]].

\bibitem{Guo:2024pvv}
B.~Guo, S.~D.~Hampton and N.~P.~Warner,
``Inscribing geodesic circles on the face of the superstratum,''
JHEP \textbf{05}, 224 (2024)
[arXiv:2401.17366 [hep-th]].

\bibitem{Dodelson:2020lal}
M.~Dodelson and H.~Ooguri,
``Singularities of thermal correlators at strong coupling,''
Phys. Rev. D \textbf{103}, no.6, 066018 (2021)
[arXiv:2010.09734 [hep-th]].


\bibitem{Horowitz22}
G.~T.~Horowitz, M.~Kolanowski and J.~E.~Santos,
``Almost all extremal black holes in AdS are singular,''
JHEP \textbf{01}, 162 (2023)
[arXiv:2210.02473 [hep-th]].

\bibitem{Horowitz23}
G.~T.~Horowitz, M.~Kolanowski, G.~N.~Remmen and J.~E.~Santos,
``Extremal Kerr Black Holes as Amplifiers of New Physics,''
Phys. Rev. Lett. \textbf{131}, no.9, 091402 (2023)
[arXiv:2303.07358 [hep-th]].

\bibitem{Horowitz24}
G.~T.~Horowitz, M.~Kolanowski, G.~N.~Remmen and J.~E.~Santos,
``Sudden breakdown of effective field theory near cool Kerr-Newman black holes,''
JHEP \textbf{05}, 122 (2024)
[arXiv:2403.00051 [hep-th]].

\bibitem{Vafa05}
C.~Vafa,
``The String landscape and the swampland,''
[arXiv:hep-th/0509212 [hep-th]].

\bibitem{Eran}
E.~Palti,
``The Swampland: Introduction and Review,''
Fortsch. Phys. \textbf{67}, no.6, 1900037 (2019)
[arXiv:1903.06239 [hep-th]].

\bibitem{Horowitz91}
G.~T.~Horowitz and A.~Strominger,
``Black strings and P-branes,''
Nucl. Phys. B \textbf{360}, 197-209 (1991)

\bibitem{Itzhaki98}
N.~Itzhaki, J.~M.~Maldacena, J.~Sonnenschein and S.~Yankielowicz,
``Supergravity and the large N limit of theories with sixteen supercharges,''
Phys. Rev. D \textbf{58}, 046004 (1998)
[arXiv:hep-th/9802042 [hep-th]].


\bibitem{Reall}
H.~S.~Reall, ``Part 3 General Relativity'', (2012)
\href{https://www.damtp.cam.ac.uk/user/hsr1000/lecturenotes_2012.pdf}{Link}

\bibitem{Emil1984}
D.~J.~Gross, J.~A.~Harvey, E.~J.~Martinec and R.~Rohm,
``The Heterotic String,''
Phys. Rev. Lett. \textbf{54}, 502-505 (1985)

\bibitem{Emil1985}
D.~J.~Gross, J.~A.~Harvey, E.~J.~Martinec and R.~Rohm,
``Heterotic String Theory. 1. The Free Heterotic String,''
Nucl. Phys. B \textbf{256}, 253 (1985)

\bibitem{Emil1985b}
D.~J.~Gross, J.~A.~Harvey, E.~J.~Martinec and R.~Rohm,
``Heterotic String Theory. 2. The Interacting Heterotic String,''
Nucl. Phys. B \textbf{267}, 75-124 (1986)


\bibitem{Gross1986}
D.~J.~Gross and J.~H.~Sloan,
``The Quartic Effective Action for the Heterotic String,''
Nucl. Phys. B \textbf{291}, 41-89 (1987)

\bibitem{Myers1987}
R.~C.~Myers,
``Superstring Gravity and Black Holes,''
Nucl. Phys. B \textbf{289}, 701-716 (1987)

\bibitem{Gross1986b}
D.~J.~Gross and E.~Witten,
``Superstring Modifications of Einstein's Equations,''
Nucl. Phys. B \textbf{277}, 1 (1986)

\bibitem{Yiming}
Y.~Chen,
``Revisiting $R^4$ higher curvature corrections to black holes,''
[arXiv:2107.01533 [hep-th]].



\bibitem{Emil93}
A.~E.~Lawrence and E.~J.~Martinec,
``Black hole evaporation along macroscopic strings,''
Phys. Rev. D \textbf{50}, 2680-2691 (1994)
[arXiv:hep-th/9312127 [hep-th]].

\bibitem{Horowitz1997b}
G.~T.~Horowitz and S.~F.~Ross,
``Properties of naked black holes,''
Phys. Rev. D \textbf{57}, 1098-1107 (1998)
[arXiv:hep-th/9709050 [hep-th]].



\bibitem{Blau} 

M. Blau, ``Plane waves and Penrose limits,'' http://www.blau.itp.unibe.ch/lecturesPP.pdf.


\bibitem{Blau2004}
M.~Blau, M.~Borunda, M.~O'Loughlin and G.~Papadopoulos,
``The Universality of Penrose limits near space-time singularities,''
JHEP \textbf{07}, 068 (2004)
[arXiv:hep-th/0403252 [hep-th]].

\bibitem{Dark}
M.~Montero, C.~Vafa and I.~Valenzuela,
``The Dark Dimension and the Swampland,''
JHEP \textbf{02}, 022 (2023)
[arXiv:2205.12293 [hep-th]].



\bibitem{Poisson:1988wc}
E.~Poisson and W.~Israel,
``Structure of the Black Hole Nucleus,''
Class. Quant. Grav. \textbf{5}, L201-L205 (1988)

\bibitem{Frolov:1988vj}
V.~P.~Frolov, M.~A.~Markov and V.~F.~Mukhanov,
``Black Holes as Possible Sources of Closed and Semiclosed Worlds,''
Phys. Rev. D \textbf{41}, 383 (1990)


\bibitem{Frolov:1989pf}
V.~P.~Frolov, M.~A.~Markov and V.~F.~Mukhanov,
``THROUGH A BLACK HOLE INTO A NEW UNIVERSE?,''
Phys. Lett. B \textbf{216}, 272-276 (1989)

\bibitem{Poisson:1990eh}
E.~Poisson and W.~Israel,
``Internal structure of black holes,''
Phys. Rev. D \textbf{41}, 1796-1809 (1990)

\bibitem{LimaJunior:2020fhs}
H.~C.~D.~Lima, Junior, L.~C.~B.~Crispino and A.~Higuchi,
``On-axis tidal forces in Kerr spacetime,''
Eur. Phys. J. Plus \textbf{135}, no.3, 334 (2020)
[arXiv:2003.09506 [gr-qc]].

\bibitem{Reynolds:2013qqa}
C.~S.~Reynolds,
``Measuring Black Hole Spin using X-ray Reflection Spectroscopy,''
Space Sci. Rev. \textbf{183}, no.1-4, 277-294 (2014)
[arXiv:1302.3260 [astro-ph.HE]].

\bibitem{Crispino:2016pnv}
L.~C.~B.~Crispino, A.~Higuchi, L.~A.~Oliveira and E.~S.~de Oliveira,
``Tidal forces in Reissner\textendash{}Nordstr\"om spacetimes,''
Eur. Phys. J. C \textbf{76}, no.3, 168 (2016)
[arXiv:1602.07232 [gr-qc]].


\bibitem{Penrose} 
R. Penrose, Singularities of spacetime, in Theoretical principles in astrophysics and relativity
(W. R. N.R. Liebowitz and P.O.Vandervoort, eds.), pp. 217–243. Chicago University Press,
1978.
[11] V. Card

\bibitem{BFSS}
T.~Banks, W.~Fischler, S.~H.~Shenker and L.~Susskind,
``M theory as a matrix model: A conjecture,''
Phys. Rev. D \textbf{55}, 5112-5128 (1997)
[arXiv:hep-th/9610043 [hep-th]].

\bibitem{Xi13}
Y.~H.~Lin, S.~H.~Shao, Y.~Wang and X.~Yin,
``A Low Temperature Expansion for Matrix Quantum Mechanics,''
JHEP \textbf{05}, 136 (2015)
[arXiv:1304.1593 [hep-th]].


\bibitem{Polchinski99}
J.~Polchinski,
``M theory and the light cone,''
Prog. Theor. Phys. Suppl. \textbf{134}, 158-170 (1999)
[arXiv:hep-th/9903165 [hep-th]].


\bibitem{Horowitz1997}
G.~T.~Horowitz and S.~F.~Ross,
``Naked black holes,''
Phys. Rev. D \textbf{56}, 2180-2187 (1997)
[arXiv:hep-th/9704058 [hep-th]].



\bibitem{Emil98a}
M.~Li, E.~J.~Martinec and V.~Sahakian,
``Black holes and the SYM phase diagram,''
Phys. Rev. D \textbf{59}, 044035 (1999)
[arXiv:hep-th/9809061 [hep-th]].

\bibitem{Emil98b}
E.~J.~Martinec and V.~Sahakian,
``Black holes and the superYang-Mills phase diagram. 2.,''
Phys. Rev. D \textbf{59}, 124005 (1999)
[arXiv:hep-th/9810224 [hep-th]].



\bibitem{Non}
C.~Akers, N.~Engelhardt, D.~Harlow, G.~Penington and S.~Vardhan,
``The black hole interior from non-isometric codes and complexity,''
[arXiv:2207.06536 [hep-th]].

\bibitem{Strings2024}
D.~L.~Jafferis, C.~Akers,  ``Black hole interiors'' discussion session in the Strings 2024 conference, CERN, June 6th 2024 ~ \href{https://lecturemedia.cern.ch/2024/1284995c161/}{Link}.




\bibitem{Gross87}
D.~J.~Gross and P.~F.~Mende,
``String Theory Beyond the Planck Scale,''
Nucl. Phys. B \textbf{303}, 407-454 (1988)

\bibitem{Mathur:2008kg}
S.~D.~Mathur,
``Tunneling into fuzzball states,''
Gen. Rel. Grav. \textbf{42}, 113-118 (2010)
[arXiv:0805.3716 [hep-th]].

\bibitem{Horowitz93}
G.~T.~Horowitz and D.~L.~Welch,
``Exact three-dimensional black holes in string theory,''
Phys. Rev. Lett. \textbf{71}, 328-331 (1993)
[arXiv:hep-th/9302126 [hep-th]].

\bibitem{Natsuume:1996ij}
M.~Natsuume and Y.~Satoh,
``String theory on three-dimensional black holes,''
Int. J. Mod. Phys. A \textbf{13}, 1229-1262 (1998)
[arXiv:hep-th/9611041 [hep-th]].

\bibitem{Giveon:2005mi}
A.~Giveon, D.~Kutasov, E.~Rabinovici and A.~Sever,
``Phases of quantum gravity in AdS(3) and linear dilaton backgrounds,''
Nucl. Phys. B \textbf{719}, 3-34 (2005)
[arXiv:hep-th/0503121 [hep-th]].

\bibitem{Bars:1990rb}
I.~Bars and D.~Nemeschansky,
``String Propagation in Backgrounds With Curved Space-time,''
Nucl. Phys. B \textbf{348}, 89-107 (1991)
\bibitem{Witten91}
E.~Witten,
``On string theory and black holes,''
Phys. Rev. D \textbf{44}, 314-324 (1991)

\bibitem{Dijkgraaf:1991ba}
R.~Dijkgraaf, H.~L.~Verlinde and E.~P.~Verlinde,
``String propagation in a black hole geometry,''
Nucl. Phys. B \textbf{371}, 269-314 (1992)

\bibitem{Maldacena:1997cg}
J.~M.~Maldacena and A.~Strominger,
``Semiclassical decay of near extremal five-branes,''
JHEP \textbf{12}, 008 (1997)
[arXiv:hep-th/9710014 [hep-th]].




\end{thebibliography}
\end{document}